\documentclass[twocolumn,aps,superscriptaddress,preprintnumbers]{revtex4}

\usepackage{amsmath}
\usepackage{graphicx}
\usepackage{amssymb}
\usepackage{epstopdf}
\usepackage{braket}
\usepackage{bm} 
\usepackage{longtable}
\usepackage{multirow}
\usepackage{enumerate}

\usepackage{hyperref}
\hypersetup{
    colorlinks=true,       
    linkcolor=blue,        
    citecolor=blue,        
    filecolor=blue,        
    urlcolor=blue,         
    runcolor=blue
}

\usepackage{xcolor}

\begin{document}

\title{Multi-level quantum Rabi model for anharmonic vibrational polaritons}

\author{Federico Hern\'andez}
\affiliation{Department of Physics, Universidad de Santiago de Chile, Av. Ecuador 3493, Santiago, Chile.}

\author{Felipe Herrera}
\email{felipe.herrera.u@usach.cl}
\affiliation{Department of Physics, Universidad de Santiago de Chile, Av. Ecuador 3493, Santiago, Chile.}
\affiliation{Millennium Institute for Research in Optics MIRO, Chile.}

\date{\today}            

\begin{abstract}
We propose a cavity QED approach to describe light-matter interaction between an individual anharmonic molecular vibration and an infrared cavity field. Starting from a generic Morse oscillator with quantized nuclear motion, we derive a multi-level quantum Rabi model to study vibrational polaritons beyond the rotating-wave approximation. We analyze the spectrum of vibrational polaritons in detail and compare with available experiments. For high excitation energies, the spectrum exhibits a dense manifold of true and avoided level crossings as the light-matter coupling strength and cavity frequency are tuned. These crossings are governed by a pseudo parity selection rule imposed by the cavity field. We also analyze polariton eigenstates in nuclear coordinate space. We show that the bond length of a vibrational polariton at a given energy is never greater than the bond length of a bare Morse oscillator with the same energy. This type of  bond hardening of vibrational polaritons occurs at the expense of the creation of virtual infrared cavity photons, and may have implications in chemical reactivity. 
\end{abstract}

\maketitle


Cavity quantum electrodynamics (QED) has been intensely studied for the development of quantum technology over the last decade \cite{Kimble2008,OBrien2009}. Precision experiments under carefully controlled conditions have been implemented to reach the regime where quantum optical effects become relevant for applications \cite{Mabuchi2002,Blais2004,RMiller2005}. Chemical systems and molecular materials at ambient conditions for long have been considered to be unnecessarily complex and uncontrollable to enable useful quantum optical effects. In recent years, the demonstration of reversible modifications of chemical properties in molecular materials via strong coupling to confined light has stimulated the study of cavity QED as an emerging research direction in chemical physics \cite{Ebbesen2016}. Light-matter interaction in the strong coupling (SC) and ultrastrong coupling (USC) regimes opens the possibility of creating novel hybrid photon-molecule states whose unique properties may enable novel applications in chemistry and material science.

In the infrared regime, the coupling of an intramolecular vibration to the quantized electromagnetic vacuum of a Fabry-P\'erot cavity can lead to the formation of vibrational polaritons \cite{Canaguier-Durand2013,Long2015,Simpkins2015,Shalabney2015raman,Shalabney2015coherent,Chikkaraddy2016,Benz2016,George2016,Vergauwe2016,Thomas2016,Hertzog2017,Wang2017,Crum2018,Chervy2018,Du2019}. These hybrid light-matter states exhibit fundamentally novel properties in comparison with free-space vibrations. For instance, vibrational polaritons may enable the selective control of chemical reactions \cite{Hiura2018,Du2019,Thomas2019}, a long-standing goal in physical chemistry \cite{Gordon1999}. Strong light-matter coupling provides a reversible way of modifying reactive processes without changing the chemical composition of materials, and also modify the radiative and non-radiative dynamics of molecular vibrations \cite{delPino2015raman,Dunkelberger2016,Dunkelberger2018,Feist2018,Ribeiro2018a,Hertzog2019,Kockum2019}. 
Several recent studies on vibrational strong coupling (VSC) within the ground electronic state have shown that chemical reactions can proceed through novel pathways in comparison with free space. Under VSC, reactions may be inhibited or catalyzed, or the product branching ratios tilted \cite{Canaguier-Durand2013,Thomas2016,Hiura2018,Thomas2019}. For instance, by strongly coupling the carbon-silyl (Si-C) bond stretching vibration of 1-phenyl-2-trimethylsilylacetylene to the electromagnetic vacuum of a resonant infrared microfluidic cavity, the rate of Si-C bond breakage has been shown to decrease by a moderate factor of order unity \cite{Thomas2016}. In another recent study \cite{Hiura2018}, the catalytic effect of VSC on the hydrolysis of cyanate ions and ammonia borane, under coupling of a Fabry-P\'erot cavity witht the broad OH infrared absorption band of water was demonstrated. The measured increase on the reaction rate constant by two orders of magnitude relative to free space for cyanate and by four orders of magnitude for ammonia borane, is one of the first reports of cavity-enhanced reactivity, a possibility earlier predicted for electron transfer reactions in microcavities \cite{Herrera2016}. In another study, very precise chemical control was carried out on a compound having two available silyl bond cleavage sites, Si–C and Si–O. For this system, simultaneous VSC with three spectrally distinct vibrational modes was shown to modify the reactivity landscape such that the branching ratio of Si–C and Si–O cleavage was altered, by simply tuning the cavity resonance conditions \cite{Thomas2019}.

The diverse experimental evidence on VSC represents a challenge for theoretical modeling, mainly due to the inherent complexity of potential energy landscapes of reactive species, as well as collective effects that are relevant in molecular ensembles. Despite recent theoretical progress \cite{DelPino2015,George2016,Muallem2016,Saurabh2016,Luk2017,Martinez-Martinez2018a,Ribeiro2018,Kockum2019,Flick2017,Flick2017a,Feist2018}, it remains unclear whether there is a universal mechanism for the modification of ground state chemical reactivity under VSC, or the problem is system-specific. Common explanations for experimental observations are based on a traditional chemical concepts such as changes in the potential energy surface, modifications of an activation energy, or changes in the relative energy of reactants and products. However, as we discuss throughout, under VSC it is difficult to justify the conventional physical meaning assigned to these traditional concepts.

In this work, we introduce a cavity QED approach to study anharmonic vibrational polaritons in the single-molecule limit. The method describes the molecular sub-system as an individual anharmonic polar bond with quantized nuclear motion. The vibration interacts with a vacuum field via electric dipole coupling. Discrete-variable representation (DVR) is used to describe the cavity-free anharmonic vibration, and the electric dipole interaction is treated with a cavity QED approach that includes counter-rotating terms. This total system Hamiltonian corresponds to a multi-level quantum Rabi (MLQR) model. By construction, vibrational polariton states can be analyzed  in Hilbert space and also in coordinate space. The method can be scaled to the many-molecule regime.

The rest of the article is organized as follows: In Section \ref{MO_sect}, we review the properties of Morse oscillators. The construction of the multi-level quantum Rabi model is discussed in Section \ref{QRM_sect}. In Section \ref{Spec_sect}, we describe the spectrum of vibrational polaritons arising from the model. In Section \ref{Coord_rep_sect}, we analyze the representation of vibrational polaritons in nuclear coordinate space. In Section \ref{Energy crossings}, we analyze the level crossings that occur in the excited polariton manifold. Section \ref{Self_E_sect} discusses the effect of the dipole self-energy term in the multipolar Hamiltonian. We conclude and discuss future developments in Section \ref{Discussion and Outlook}.

\section{Morse oscillator} \label{MO_sect}

\begin{figure}[t]
\includegraphics[width=0.5\textwidth]{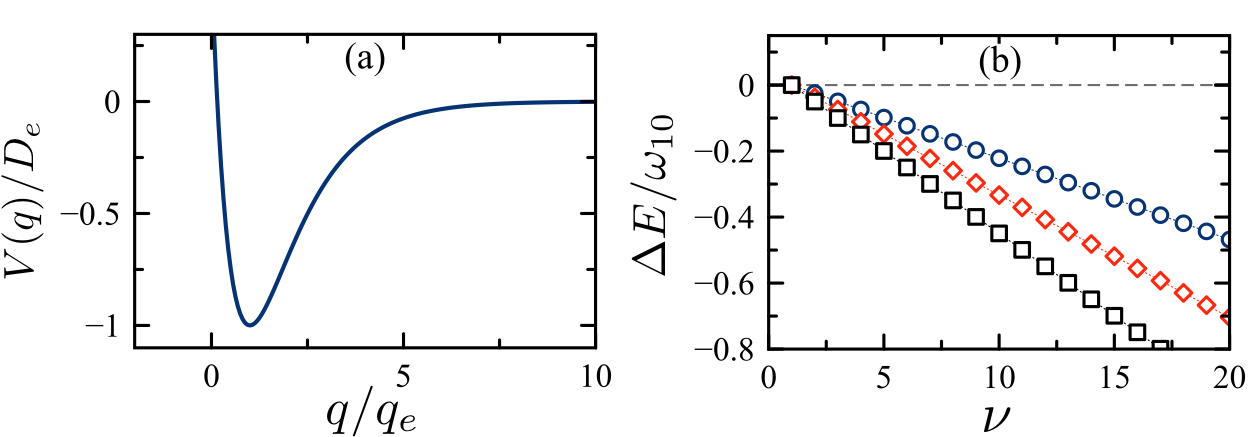}
\caption{(a) Morse potential $V(q)$, in units of the classical dissociation energy $D_e$. We use $D_e = 12$, $q_e = 4$, $a=0.204$ (dimensionless) to generate 24 bound states. (b) Anharmonicity of the energy spacing between adjacent Morse eigenstates  $\Delta E\equiv \omega_{\nu,\nu-1}-\omega_{10}$. We use $D_e=12.0$, $q_e = 4.0$ for all points, with $\mu=3$ and $a=0.204$ (circles), $\mu=1$ and $a=0.175$ (diamonds), and $\mu=1$ and $a=0.233$ (squares). The dashed line is the harmonic oscillator limit $(\Delta E = 0)$.}
\label{fig:morse potential}
\end{figure}

We model the nuclear motion of an anharmonic polar vibration (e.g. carbonyl) with a Morse potential \cite{Morse1929}
\begin{equation}
V(q) = D_e\left(1-{\rm exp}[-a(q-q_e)]\right)^2,
\label{eq:MO_pot}
\end{equation}
where $D_e$ is the classical dissociation energy (without zero point motion), $q_e$ is the equilibrium bond length, and $a$ is a parameter. The vibrational Schr\"{o}dinger equation with a Morse potential can be solved analytically in terms of associated Laguerre polynomials \cite{Morse1929,Vasan1983}, or numerically using grid-based methods \cite{Light2007}. We use DVR on a uniform grid with Fourier basis functions \cite{Colbert1992} to obtain the vibrational wavefunctions and eigenvalues of the Morse potential. For the dimensionless parameters $D_e=12.0$, $q_e = 4.0$, and $a=0.2041$, the corresponding potential is shown in Fig. \ref{fig:morse potential}a. For the dimensionless mass $\mu=1$, this potential has 24 bound states and is used throughout. In the Supplemental Material, we show that our conclusions do not vary qualitatively for other Morse potential parameters.

As we explain below, the degree of anharmonicity in the potential has a profound effect in the behaviour of vibrational polaritons. For the Morse potential, the anharmonicity can be easily tuned by changing the parameters $a$ and $\mu$, for fixed binding energy $D_e$. The relation between these parameters and the degree of anharmonicity can be understood from the exact eigenvalues of the Morse Hamiltonian \cite{Morse1929}
\begin{equation}\label{eq:MO_eigenenergies}
E_{\nu} =-D_e+a \hbar\sqrt{\frac{2D_e}{\mu}}\,(\nu+1/2)-\frac{a^2\hbar^2}{2D_e \mu} (\nu+1/2)^2
\end{equation}
where $\nu$ is the vibrational quantum number. By comparing this expression with the Dunham expansion \cite{Demtroder-book} 
\begin{eqnarray}\label{eq:Dunham_expansion}
E_{\nu} &=&Y_{00} +\omega_{0}(\nu+1/2)-\omega_0\chi_{e} (\nu+1/2)^{2}+\ldots 
\end{eqnarray}
where $\omega_{0}$ is the vibrational frequency, the anharmonic coefficient $\chi_e$ can be written as
\begin{equation}\label{eq:chi_e}
\chi_{e} = \frac{\pi\hbar^2 a}{(2\mu)^{1/2} D_e ^{3/2} }.
\end{equation}
Vibrations with lower $\mu$ and higher $a$ therefore have stronger spectral anharmonicity, for fixed dissociation energy. We illustrate this dependence in Fig. \ref{fig:morse potential}b, where the change in the vibrational energy level spacing relative to the fundamental frequency $\omega_{10}$ for $\nu=1\leftarrow \nu=0$ is shown for different values of $\mu$ and $a$. The level spacing between adjacent vibrational states can be significantly smaller than the harmonic oscillator value $\omega_{10}$, even for relatively low values of $\nu$. 

\begin{figure*}
\includegraphics[width=1\textwidth]{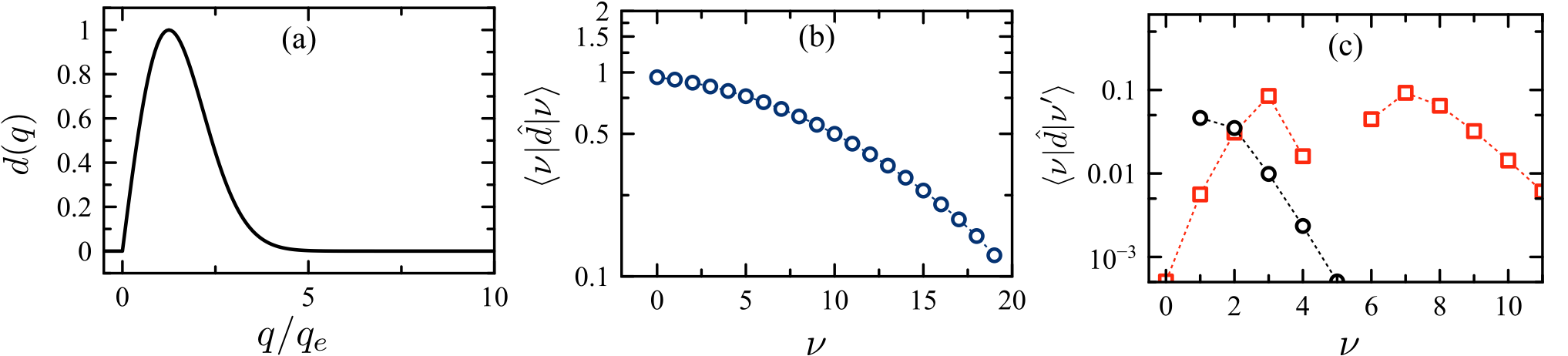}
\caption{(a) Rayleigh distribution model for the electric dipole function of a polar bond $d(q)$ normalized to its maximum value. (b) Permanent dipole moment matrix elements $\bra{\nu}d_e(q)\ket{\nu}$ as a function of the vibrational quantum number $\nu$. (c) Dipole matrix elements $\bra{\nu}d_e(q)\ket{\nu'}$ as a function of $\nu$, for $\nu'=0$ (circles) and $\nu' = 5$ (squares). Dipole matrix elements are normalized to the maximum of $d(q)$. The Morse parameters used are $D_e=12.0$, $q_e = 4.0$, $a=0.204$ and $\mu=1$.}
\label{fig:dipole_moments}
\end{figure*}

The electronic wavefunction determines the contribution to the molecular dipole moment of the electron charge distribution, which in the Born-Oppenheimer approximation is a parametric function $d(\{\mathbf{q}\})$ of all nuclear coordinates $\{\mathbf{q}\}$. In general, the dipole function $d(\{\mathbf{q}\})$ can be obtained using {\it ab-initio} quantum chemistry for simple molecular species. Since we are interested in understanding universal features of anharmonic vibrational polaritons, we adopt a model functional form $d(q)$ that captures the correct physical behaviour of a one-dimensional polar bond. The function must be: ({\it i}) continuous over the entire range of $q$; ({\it ii}) have a maximum at some value of $q$, not necessarily the equilibrium distance; ({\it iii}) asymptotically vanish as the neutral bond dissociates into neutral species. These requirements are satisfied by a Rayleigh distribution function of the form
\begin{equation}
d(q) = (q+c_0)\exp{[-q^2 / 2\sigma^2]},
\label{eq:Rayleigh_eq}
\end{equation}
which we show in Fig. \ref{fig:dipole_moments}a for $c_0=0$ and $\sigma/q_e=1.25$. These parameters are used throughout. $\sigma$ is the coordinate at which the electronic dipole moment is greater. The parameter $c_0$ is the dipole moment for $q/q_e\ll 1$. In the Supplemental Material, we show that our results and conclusions do not qualitatively vary for different choices of $\sigma$ and $c_{0}$.

In order to describe light-matter coupling properly, we not only need a reasonable description of the electric dipole moment near the equilibrium distance $q_e$, but also in the long range up to the dissociation threshold. This is because strong light-matter coupling in a cavity can strongly admix several vibrational eigenstates with high $\nu$. Furthermore, we are interested in studying how highly exited polaritons behave near the energy dissociation threshold of the free-space molecular sub-system. Therefore, dipole matrix elements between all bound and unbound states of the Morse potential must be  accurately estimated. In Figs. \ref{fig:dipole_moments}b and \ref{fig:dipole_moments}c we show the scaling with $\nu$ of the diagonal and off-diagonal vibrational dipole matrix elements. The permanent dipole moments $\bra{\nu}\hat{d}(q)\ket{\nu}$ decrease with $\nu$ (panel \ref{fig:dipole_moments}b), as expected from the behaviour of $d(q)$ on a neutral molecular system. This trend also holds for Morse oscillators with different $a$ and $\mu$ parameters. The higher the oscillator's mass, the lower the rate of decrease. Fig. \ref{fig:dipole_moments}c shows that for a fixed vibrational eigenstate $\ket{\nu'}$, the transition dipole moments with neighbouring states $\ket{\nu}$ ($\nu\neq \nu'$) are not negligible, and must be taken into account in the light-matter coupling. 

In linear infrared absorption of high-frequency modes  (e.g., $\omega_{10}\approx 200$ meV for carbonyl), only the ground vibrational level $\nu=0$ is populated at room temperature ($kT/\hbar \omega_{10}\ll 1$). The oscillator strength of the fundamental absorption peak ($\nu=0\rightarrow 1$) and its overtones ($|\Delta \nu| \geq 2$) are thus proportional to  $|\bra{\nu}\hat{d}(q)\ket{0}|^2$ with $\nu\geq 1$. Figure \ref{fig:dipole_moments}c (circles) captures the typical IR absorption pattern of decreasing overtone strength for higher
$\Delta\nu$ \cite{Jakob1998}. 
This qualitatively correct behaviour validates the dipole model function $d(q)$ in Eq. (\ref{eq:Rayleigh_eq}). 

Using strong infrared laser pulses, it is possible to prepare vibrational modes with high quantum numbers $\nu\gg 1$, even when $kT/\hbar \omega_{10}\ll 1$. This off-resonant driving is known as vibrational ladder climbing, and has been used in nonlinear spectroscopic measurements \cite{Marcus2006}. Vibrational ladder climbing is determined by the matrix elements
$\bra{\nu}\hat{d}\ket{\nu'}$ with $\nu'\neq \nu\geq 1$, corresponding to dipole transitions between overtones. Fig. \ref{fig:dipole_moments}c (squares) shows that these high-$\nu$ matrix elements can be as strong as the first overtones of the fundamental transition (circles), over a range of neighbouring levels with $|\nu-\nu'| \leq 4$, for our choice of $d(q)$. 
We show below that ignoring dipole couplings between high-$\nu$ overtones fails to describe the rich and complex physics of the excited polariton manifold up to the dissociation threshold. Excited polariton levels can be expected to be relevant in the description of nonlinear cavity transmission signals, chemical reactions, and heat transport.

\section{Multi-level quantum Rabi model} \label{QRM_sect}

We derive the total Hamiltonian for the molecule-cavity system starting using the Power-Zineau-Wolley (PZW) multipolar formulation of light-matter interaction \cite{Craig-Thirunamachandran-book}. The PZW frame is equivalent to minimal-coupling by a unitary transformation that eliminates the vector potential $\mathbf{A}(\mathbf{x})$ from the Hamiltonian \cite{Craig-Thirunamachandran-book}. We divide the total Hamiltonian $\hat{\mathcal{H}}$ in the three terms of the form $\mathcal{\hat{H}} = \hat{H}_{\mathrm{M}} + \hat{H}_{\mathrm{C}} + \hat{H}_{\mathrm{LM}}$. The molecular part is given by
\begin{equation}\label{eq:Hmol}
\hat{H}_{\rm M} = \hat{H}_{\rm el} + \hat{H}_{\rm vib} + \hat{H}_{\rm rot} + \int d\mathbf{x}\; |\mathbf{P}(\mathbf{x})|^{2},
\end{equation}
where the first three terms represent the electronic, vibrational and rotational contributions, respectively. The last term corresponds to the dipole self-energy, with $\mathbf{P}(\mathbf{x})$ being the macroscopic polarization density. 

The free cavity Hamiltonian  $\hat H_{\rm C}$ is given by 
\begin{eqnarray}\label{eq:Hcav}
\hat{H}_{\rm C} &=& \frac{1}{2} \int d\mathbf{x}\left( |\mathbf{D}(\mathbf{x})|^{2} + \frac{1}{\mu_{0}}|\mathbf{H}(\mathbf{x})|^{2}\right)  \nonumber\\
&= & \sum_{\xi} \hbar\omega_{\xi} \left(\hat{a}_{\xi}^{\dag} \hat{a}_{\xi} + 1/2\right),
\end{eqnarray}
where $\mathbf{D}(\mathbf{x})$ and $\mathbf{H}(\mathbf{x})$ are the macroscopic displacement and magnetic fields, respectively. $\mu_{0}$ is the magnetic permeability. In the second line, we imposed canonical field quantization into a set of normal modes with continuum label $\xi$,  frequencies $\omega_\xi$, and annihilation operators $\hat a_\xi$. Light-matter interaction in the PZW frame, ignoring magnetic moments, is given by \cite{Craig-Thirunamachandran-book}
\begin{equation}\label{eq:H-LM}
\hat{H}_{\rm LM} = \int d\mathbf{x}\; \mathbf{P}(\mathbf{x}) \cdot \mathbf{D}(\mathbf{x}). 
\end{equation}

We consider a non-rotating polar bond and therefore set $\hat{H}_{\rm rot}=0$. The solutions of the electronic Hamiltonian $\hat{H}_{\rm el}$ are assumed to be known within the Born-Oppenheimer approximation, such that they give the dipole function $d(q)$. The self-energy term in Eq. (\ref{eq:Hmol}) will be shown to produce a state-dependent vibrational shift that does not qualitatively affect the polariton spectrum and eigenstates, and can be ignored to simplify the analysis. In Section \ref{Self_E_sect} we put the dipole self-energy back into the Hamiltonian and discuss its effect on the polariton spectrum. We  adopt a point-dipole approximation for the polarization density, i.e. $\mathbf{P}(\mathbf{x})=\mathbf{d}\,\delta(\mathbf{x}-\mathbf{x}_0)$, where $\mathbf{d}$ is the electric dipole vector and $\mathbf{x}_0$ is the location of the molecule.

We use a single-mode approximation for the cavity Hamiltonian in Eq. (\ref{eq:Hcav}) by setting $\omega_\xi\equiv \omega_c$ for all $\xi$ and defining the effective field operators $\hat a =\sum_\xi \hat a_\xi$ (up to a normalization constant). This simplification is justified in Fabry-P\'erot cavities with a large free-spectral range (FSR $\sim 300-500$ cm$^{-1}$ \cite{Simpkins2015}), and low transmission  linewidths (FWHM $\sim 10-40$ cm$^{-1}$ \cite{Simpkins2015}). In this approximation, the intracavity displacement field operator can be approximated by $\hat{\mathbf{D}} \approx \mathcal{E}_{0}(\hat{a}+\hat{a}^{\dag})$, where $\mathcal{E}_0$ can be considered as the amplitude of the vacuum field fluctuations, or the electric field per photon (ignoring vectorial character). $\mathcal{E}_0$ scales as $1/\sqrt{V_m}$ with the effective cavity mode volume $V_m$ \cite{Novotny-hecht-book}. We thus write the light-matter interaction term as
\begin{equation} \label{eq:H-LM2}
\hat{H}_{\rm LM} = \mathcal{E}_{0}\,(\hat{d}_{+}+\hat{d}_{-})\otimes (\hat{a}+\hat{a}^{\dag}),
\end{equation}
where the up-transition operator $\hat d_+$ projected into the vibrational energy basis $\ket{\nu}$ is given by
\begin{equation}
\hat{d}_{+} = \sum_{\nu,\nu'>\nu}\bra{\nu'}{d}(q)\ket{\nu}\ket{\nu'}\bra{\nu},
\label{eq:trans_dipole}
\end{equation}
with $\hat{d}_{-}=(\hat{d}_{+})^{\dag}$. By combining Eqs. (\ref{eq:Hmol}) without self-energy, Eq. (\ref{eq:Hcav}) in the single-mode approximation, and Eq. (\ref{eq:H-LM2}), we can arrive at the total system Hamiltonian 
\begin{eqnarray}\label{eq:Rabi_Ham}
\mathcal{\hat{H}} &=& \omega_{c} \,\hat{a}^{\dag} \hat{a} +\sum_{\nu}\omega_{\nu}\ket{\nu}\bra{\nu}  \\
&&+\sum_{\nu}\sum_{\nu'>\nu} g_{\nu'\nu}(\ket{\nu'}\bra{\nu}+\ket{\nu}\bra{\nu'})(\hat{a}+\hat{a}^{\dag})\nonumber
\end{eqnarray}
where $\omega_\nu$ is the energy of the vibrational eigenstate $\ket{\nu}$, and  $g_{\nu'\nu}=\mathcal{E}_{0}\bra{\nu'}d(q)\ket{\nu}$ for $\nu'>\nu$ is a state-dependent  Rabi frequency. The zero of energy is defined by the energy of the vibrational ground state ($\nu=0$) in the cavity vacuum. Equation (\ref{eq:Rabi_Ham}) corresponds to a multi-level quantum Rabi (MLQR) model, which reduces to the quantum Rabi model for a two-level system \cite{Werlang2008,Braak2011,Wolf2013}, when the vibrational space is truncated to $\nu=0,1$, and the energy reference rescaled.

The vacuum field amplitude $\mathcal{E}_0$ is considered here as a tunable parameter that determines the light-matter coupling strength. In a cavity with small mode volume, the mode amplitude $\mathcal{E}_0$ can be large and tunable by fabrication \cite{Chikkaraddy2016}. Moreover, the cavity detuning $\Delta\equiv \omega_c-\omega_{10}$ is another energy scale that can be tuned by fabrication.

For convenience, we define the state-independent Rabi frequency 
\begin{equation}
g\equiv g_{10}=E_0\,\langle 1|  d(q) |0 \rangle.
\label{eq:g_parameter}
\end{equation}
Although we use the single parameter $g$ to quantify light-matter coupling strength throughout, we emphasize that dipole transitions $\nu \leftrightarrow \nu'$ in Eq. (\ref{eq:Rabi_Ham}) have in general different coupling strengths.

\section{Spectrum of vibrational polaritons} \label{Spec_sect}

\begin{figure}[t]
\includegraphics[width=0.5\textwidth]{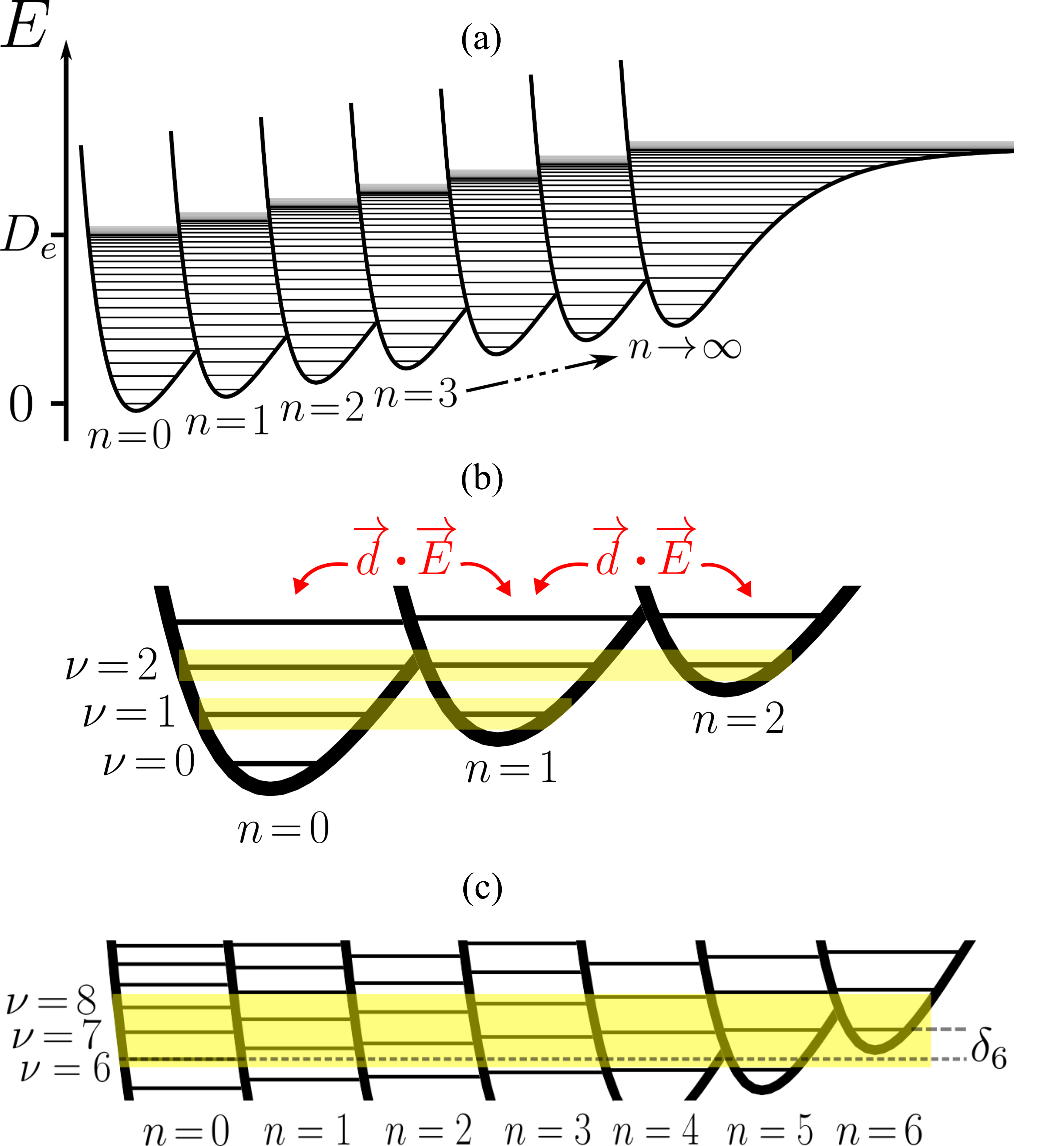}

\caption{(a) Illustration of resonant light-matter  coupling between a Morse oscillator with dissociation energy $D_e$ and a quantized cavity field with photon number $n$ (unbound). Each Morse potential corresponds to the uncoupled subspace $\ket{\nu}\ket{n}$. (b) Low energy couplings involving the subspace $\mathcal{S}_1=\{\ket{1}\ket{0}, \ket{0}\ket{1}\}$ at $E\approx \omega_{10}$, and $\mathcal{S}_2=\{\ket{2}\ket{0},\ket{1}\ket{1}, \ket{0}\ket{2}\}$ at $E\approx 2\omega_{10}$. Dipole coupling within $\mathcal{S}_1$ leads to the formation of the lower and upper polaritons, and coupling within $\mathcal{S}_2$ to the formation of a polariton triplet. (c) High energy couplings involving $\mathcal{S}_6$ at $E\approx 6\omega_{10}$. State $\ket{6}\ket{0}$ is red shifted with respect to $\ket{0}\ket{6}$ by $\delta_6$, for $\omega_{10}=\omega_c$. The highlighted levels can strongly admix.}
\label{fig:Coupling}
\end{figure}

In order to gain some physical intuition about the  structure of vibrational polaritons, in Fig. \ref{fig:Coupling} we illustrate the light-matter coupling scheme implied by the uncoupled basis $\ket{\nu}\ket{n}$, where $\ket{n}$ is a cavity Fock state. 
We can associate a complete vibrational manifold $\{\ket{\nu}\,; \nu=0,1,2,\ldots\}$ to every Fock state of the cavity $\ket{n}$. The ground level in each vibrational manifold ($\nu=0$) has energy $n\omega_c$ in the Fock state $\ket{n}$, and the dissociation energy $E_\infty$ becomes
\begin{equation}\label{eq:Threshold}
 E_\infty=D_e+n\omega_c, 
\end{equation}
Only in the cavity vacuum ($n=0$), the bond dissociation energy coincides with the value expected for a Morse oscillator in free space. In general, the energy required to break a chemical bond depends on quantum state of the cavity field. 

Vibrational manifolds with different Fock states can couple each other via the light-matter term in Eq. (\ref{eq:H-LM2}). Since parity is broken for vibrational states due to anharmonicity, the only quasi selection rule that holds is $\Delta n=\pm 1$, because the free cavity Hamiltonian $\hat H_{\rm C}$ commutes with parity. Therefore, vibrational states $\ket{\nu}$ and $\ket{\nu'}$ that differ by one photon number can admix due to light-matter coupling. Because of anharmonicity, admixing of vibrational states with $|\nu-\nu'| \geq 1$ is allowed. The amount of admixing that can occur between vibrational eigenstates in different manifolds is ultimately determined by the electric dipole function $d(q)$. 

The number bare states $\ket{\nu}\ket{n}$ that can potentially admix to form vibrational polariton eigenstates grows as the total energy increases. Figure \ref{fig:Coupling}b shows that for the lowest Fock states, resonant coupling at energy $E\approx \omega_{10}$ only involves the subspace $\mathcal{S}_1=\{\ket{1}\ket{0}, \ket{0}\ket{1}\}$ for $g/\omega_{10}\ll 1$. This coupling results in the formation of the so-called lower polariton (LP) and upper polariton (UP), which are observable in linear spectroscopy \cite{Ebbesen2016,Simpkins2015}. They can be written as 

\begin{subequations}\label{eq:LP-UP}
\begin{align}
\ket{\Psi_{1}} = \alpha \ket{0}\ket{1}- \beta \ket{1}\ket{0} \\
\ket{\Psi_{2}} = \beta \ket{0}\ket{1}+ \alpha \ket{1}\ket{0}
\end{align}
\end{subequations}
where $\ket{\Psi_{1}}$ and $\ket{\Psi_{2}}$ correspond to LP and UP, respectively. The orthonormal coefficients $\alpha$ and $\beta$ depend on $g$ and $\Delta$. $\ket{\Psi_{1}}$ and $\ket{\Psi_{2}}$ in Eq. (\ref{eq:LP-UP}) coincide with the first excitation manifold of the Jaynes-Cummings model \cite{jaynes1963}. 
Figure \ref{fig:Coupling}b also shows that for $g/\omega_{10}\ll 1$, resonant coupling at energy $E\approx 2\omega_{10}$ only involves the subspace $\mathcal{S}_2=\{\ket{2}\ket{0}, \ket{1}\ket{1}, \ket{0}\ket{2}\}$, leading to the formation of three polariton branches, as discussed below. For $g/\omega_{10}\sim 0.1$ coupling of bare states $\ket{\nu}\ket{n}$ beyond $\mathcal{S}_1$ and $\mathcal{S}_2$ is allowed by counter-rotating terms in Eq. (\ref{eq:H-LM2}).

In Fig. \ref{fig:Coupling}c, we consider the coupling between vibrational manifolds around energy $E\approx 6\,\omega_{10}$. If the molecular vibrations were harmonic, vibrational states $\ket{\nu}$ would have energy $\nu\omega_{10}$. Due to anharmonicity, vibrational levels in free space have energy 
$$\omega_{\nu}=\nu\,\omega_{10}-\delta_\nu,$$
where $\delta_\nu>0$ is the shift from a harmonic oscillator level, shown in Fig. \ref{fig:morse potential} for a Morse oscillator. For $\nu=6$, the anharmonic shift $\delta_6$ is not negligible in comparison with $\omega_{10}$, which means that for the smaller couplings $g/\omega_{10}\ll 1$, the number of bare states $\ket{\nu}\ket{n}$ that can resonantly admix is relatively limited. This resembles the role of anharmonicity in limiting the efficiency of vibrational ladder climbing using laser pulses \cite{Kuhn1999,Marcus2006}.

\begin{figure*}[t]
\includegraphics[width=\textwidth]{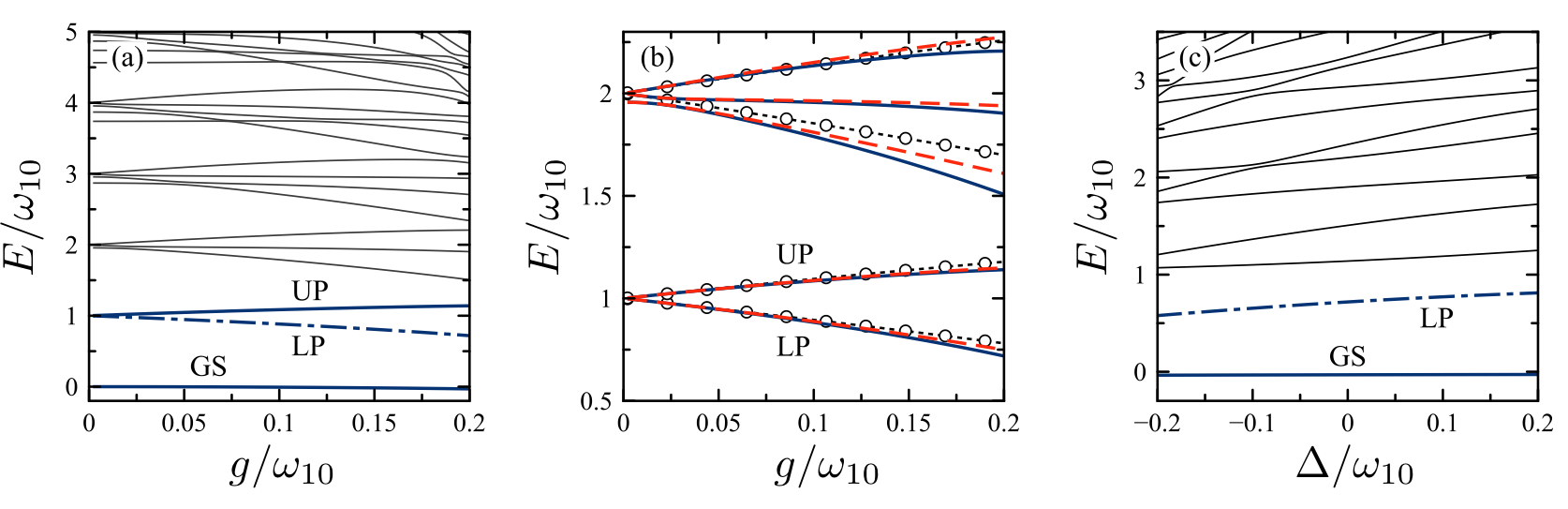}
\caption{(a) Spectrum of anharmonic vibrational polaritons as a function of the coupling strength $g/\omega_{10}$, for resonant coupling $\omega_{c} = \omega_{10}$. The ground state (GS), lower (LP) and upper (UP) polaritons are highlighted. (b) Spectrum of the lowest five excited polaritons obtained by three levels of theory: the multi-level quantum Rabi model of Eq. (\ref{eq:Rabi_Ham}) (solid line), the quantum Rabi model for a two-level vibration $\nu=\{0,1\}$ (open circles) and an anharmonic three-level quantum Rabi model with $\nu = \{0,1,2\}$ (dashed  line). (c) Vibrational polaritons spectrum as a function of the cavity detuning from the fundamental frequency $\Delta = \omega_c - \omega_{10}$. The GS and LP are highlighted. We set $g=0.2 \omega_0$. Energy is in units of $\omega_{10}$.}
\label{fig:polariton spectrum}
\end{figure*}

On the other hand, Fig. \ref{fig:Coupling}c suggests that for larger coupling ratios $g/\omega_{10}$ there is a greater number of quasi-degenerate bare states $\ket{\nu}\ket{n}$ that are energetically available to admix within an energy range $2\delta$. As the total energy increases, the density of quasi-degenerate bare states that can strongly admix within a bandwidth $\delta E$ grows. We show below that this complex coupling structure leads to a large density of true and avoided crossings in the excited polariton manifold, even for relatively low values of the coupling ratio $g/\omega_{10}$.  

In Fig. \ref{fig:polariton spectrum}, we show the spectrum of anharmonic vibrational polaritons as a function of $g$ and $\Delta$. Figure \ref{fig:polariton spectrum}a shows that the system has a unique non-degenerate ground state $\ket{\Psi_0}$ (GS). The first excited manifold features a LP-UP doublet that scales linearly with $g$ over the range of couplings considered ($R^2=1.000$ for log-log fit). However, the LP-UP splitting is not symmetric around $E=\omega_{10}$, which is the energy of the degenerate bare states $\ket{1}\ket{0}$ and $\ket{0}\ket{1}$ for $\omega_c=\omega_{10}$. Fig. \ref{fig:polariton spectrum}a shows the polariton triplet around $E=2\omega_{10}$, associated with light-matter coupling within the subspace $\mathcal{S}_2$ discussed above (see Fig. \ref{fig:Coupling}b). Multiple true and avoided crossings occur at energies $E\geq 2\omega_{10}$ over the entire range of couplings considered. The density of energy crossings grows with increasing energy. 

In Fig. \ref{fig:polariton spectrum}b we compare the energies of the lowest five excited states obtained by three levels of theory: ({\it i}) the MLQR model in Eq. (\ref{eq:Rabi_Ham}); ({\it ii}) the quantum Rabi model for a two-level vibration involving states $\nu=\{0,1\}$; ({\it iii}) a three-level quantum Rabi model with $\nu=\{0,1,2\}$ which takes into account the anharmonicy shift of the transition $\nu=1\rightarrow \nu=2$, i.e., $\omega_{21}=\omega_{10}-\delta_2$. The latter was used in Ref. \cite{Ribeiro2018} to interpret the intracavity differential absorption spectrum of W(CO$_2)_6$ \cite{Xiang2018}. By construction, the qubit model can qualitatively match the asymmetric LP-UP splitting around $E=\omega_{10}$ over the range $g/\omega_{10}\leq 0.1$, but deviations occur for larger coupling strengths. Since $g/\omega_{10}= 0.1$ is conventionally regarded as the onset of the ultrastrong coupling regime \cite{Kockum2019}, the deviations of the two-level model from MLQR for $g/\omega_{10}> 0.1$, can be attributed to the inability of the truncated two-level model to capture counter-rotating overtone couplings properly. By increasing the dimensionality of the vibrational basis by one additional state ($\nu=2$), the three-level quantum Rabi model matches better the LP-UP spectrum predicted by the MLQR model, but is  unable to correctly capture the splitting of the polariton triplet around energy $E=2\,\omega_{10}$, except for the smallest coupling ratios ($g/\omega_{10}\ll 1$).

The comparison between models in Fig. \ref{fig:polariton spectrum}b suggests that for the excited vibrational polaritons considered, the onset of ultrastrong coupling--where counter-rotating terms in the light-matter interaction becomes important--occurs at much smaller values of $g$ than those expected for a qubit, and can involve off-resonant coupling to higher vibrational levels with $\nu\geq 3$. For excited polaritons with energies $E\geq 3\,\omega_{10}$, few-level truncations of the material Hamiltonian (e.g. Ref. \cite{Ribeiro2018}) fail to capture the multiple true and avoided crossings that the Hamiltonian allows. We further discuss these excited state crossings in Section \ref{Energy crossings}.

In Fig. \ref{fig:polariton spectrum}c, we show the polariton spectrum as a function of detuning $\Delta\equiv\omega_{c}-\omega_{10}$, for $g/\omega_{10}=0.2$. Several true and avoided crossings develop in the excited manifold. When $\Delta\sim g$, the energetic ordering of the excited polaritons can change in comparison with the resonant regime ($\Delta/g\ll 1$). For example, there is an avoided crossing at $E\approx 2.1\,\omega_{10}$ near $\Delta\approx -0.1\omega_{10}$. The upper polariton (UP) also crosses with the next excited polariton level at $\Delta\approx -0.28\,\omega_{01}$. This raises concerns regarding the assignment of spectral lines in linear and nonlinear cavity transmission spectroscopy for light-matter coupling in the dispersive regime $|\Delta|/g \gtrsim 1$.

\section{Vibrational polaritons in nuclear coordinate space} \label{Coord_rep_sect}

In molecules and materials, the strength of a chemical bond is commonly associated with its vibration frequency $\omega_0$ via the relation 
\begin{equation}
    \omega_0 = \sqrt{k/\mu},
\end{equation}
where $k$ is the bond spring constant and $\mu$ is the reduced mass of the vibrating  nuclei. Stronger bonds (higher $k$) thus lead to higher vibrational frequencies. This simple argument has also been used to discuss the bonding character of vibrational polaritons under strong coupling \cite{Ebbesen2016}. In this Section, we show that the description of the bonding strength of vibrational polaritons is far more complex than the commonly used spring model suggests. 

\begin{figure}[t]
\includegraphics[width=0.48\textwidth]{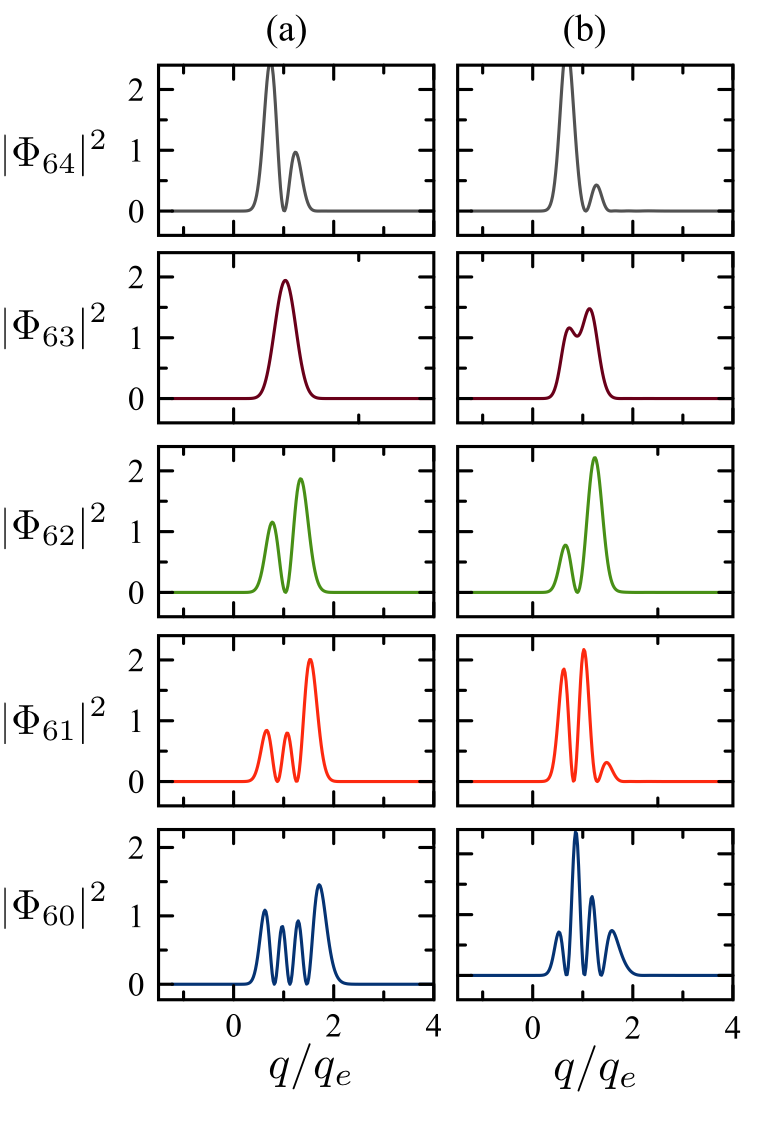}
\caption{Conditional probability densities $|\Phi_{6n}(q)|^2$ for the excited polariton eigenstate $\ket{\Psi_6}$, for coupling strengths $g/\omega_{10}=0.002$ (a) and $g/\omega_{10}=0.2$ (b). Coordinates are in units of the bare equilibrium bond length $q_e$. All densities are normalized.}
\label{fig:spacial_rep}
\end{figure}

In order to analyze vibrational polaritons in nuclear coordinate space, keeping photons in Hilbert space, let us expand the eigenstates of Eq. (\ref{eq:Rabi_Ham}) in the uncoupled basis $\{\ket{\nu}\ket{n}\}$ as
\begin{equation}\label{eq:Psi_j expansion}
    \ket{\Psi_j} = \sum_{\nu,n}c_{\nu n}^{j}\ket{\nu}\ket{n},
\end{equation}
where $c_{\nu n}^{j}$ are orthonormal coefficients associated with the $j$-th eigenstate. We can rewrite Eq. (\ref{eq:Psi_j expansion}) by combining vibrational components associated with a given photon number $n$ as 
\begin{equation}
    \ket{\Psi_j} = \sum_{n}\ket{\Phi_n^j}\ket{n},
\end{equation}
where $\ket{\Phi_n^j} = \sum_\nu c_{\nu n}^j\ket{\nu}$. The state $\ket{\Phi_n^j}$ can be interpreted as a vibrational wavepacket conditional on the cavity photon number. Its nuclear coordinate representation is simply given by the projection
\begin{equation}
    \Phi_{jn}(q)=\langle q|\Phi_{n}^j\rangle. 
\end{equation}
For concreteness, we show in Fig. \ref{fig:spacial_rep} a set of normalized conditional  probability distributions $|\Phi_{jn}(q)|^2$ with $n\leq 4$, for the excited polariton eigenstate $\ket{\Psi_6}$ under resonant light-matter coupling. Since the energy of excited polariton $\ket{\Psi_6}$ tends asymptotically to $E_6\approx 3\omega_{10}$ as $g/\omega_{10}\rightarrow 0$, one could expect the normalized probability distribution $|\Phi_{6n}(q)|^2$ to resemble the behaviour of the Morse oscillator eigenfunction with $\nu = 3$ for $g/\omega_{10}\ll 1$. Figure \ref{fig:spacial_rep} (lower panel) shows that indeed the vacuum component ($n=0$) of $\ket{\Psi_6}$ qualitatively matches the node structure of the bare Morse oscillator state  $\ket{\nu=3}$ for $g/\omega_{10}=0.002$. However, for the coupling ratio $g/\omega_{10}=0.2$, the nuclear density of the cavity vacuum $|\Phi_{60}(q)|^2$ behaves qualitatively different from a Morse eigenfunction. Similar deviations from the bare Morse behavior occurs also for nuclear components with higher photon numbers ($n\geq 1$).

\begin{figure}[t]
\includegraphics[width=0.48\textwidth]{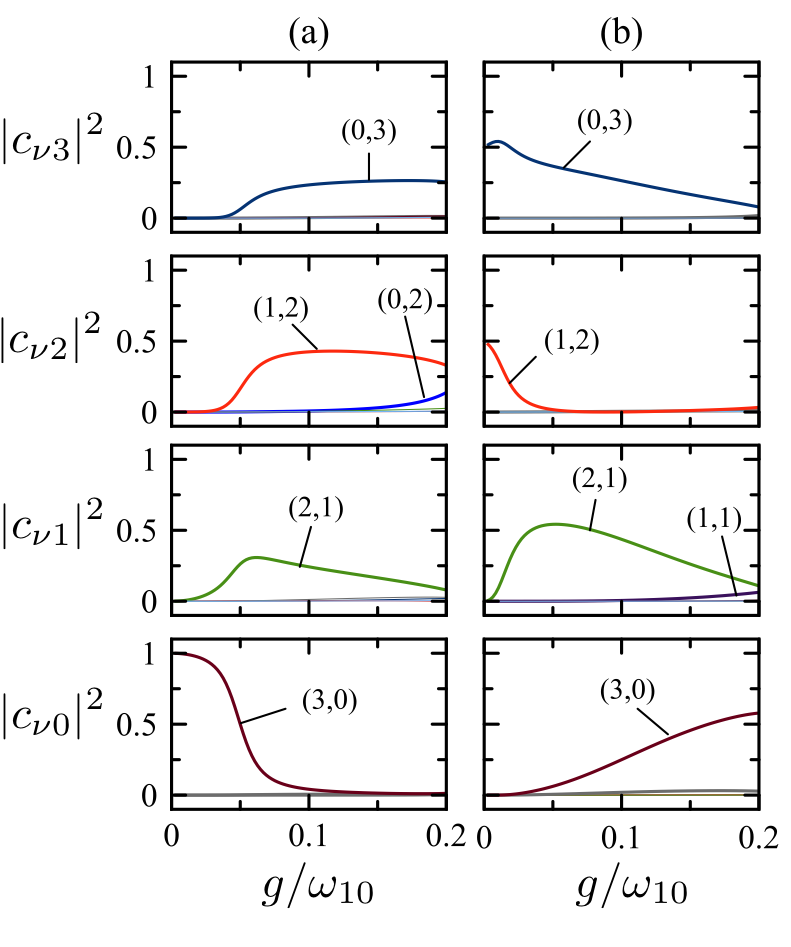}
\caption{Probability amplitudes $|c_{\nu n}|^2$ in the uncoupled basis $\ket{\nu}\ket{n}$ for excited polariton eigenstates $\ket{\Psi_6}$ (a) and $\ket{\Psi_8}$ (b), as a function the coupling strength $g/\omega_{10}$. Curves are labelled by the quantum numbers $(\nu,n)$. We set $\omega_c=\omega_{10}$.}
\label{fig:eigen_coeffs}
\end{figure}

Figure \ref{fig:spacial_rep} also shows that the nuclear densities $|\Phi_{6n}(q)|^2$ associated with  $n\geq 1$ can also approximately resemble the node pattern of a bare Morse oscillator with the appropriate number of excitations, for small values of $g/\omega_{10}$. For example, since the energy of $\ket{\Psi_6}$ tends to $E\approx 3\omega_{10}$ as $g\rightarrow 0$, its wave function should have components in the uncoupled basis $\ket{\nu}\ket{n}$ such that $\nu + n = 3$ at zero detuning ($\omega_c=\omega_{10}$). For $g/\omega_{10}=0.002$, Fig. \ref{fig:spacial_rep} shows that indeed for $n=1$ the nuclear density $|\Phi_{61}(q)|^2$ of state $\ket{\Psi_6}$ has a node structure similar to the bare Morse eigenstate $\ket{\nu=2}$, i.e., it has two nodes. The nuclear densities associated with $n=2$ and $n=3$ also seem to satisfy a conservation rule for the total number of excitations $(\nu+n)$. This rule however is broken for the $n=4$ nuclear wave packet $ \Phi_{64}(q)  $ (Fig. \ref{fig:spacial_rep}, upper panel), which has a node structure similar to the bare Morse eigenstate $\ket{\nu=1}$, corresponding to a total number of excitations $\nu+n=5$ for all values of $g/\omega_{10}$ considered.

\begin{figure}[t]
\includegraphics[width=0.48\textwidth]{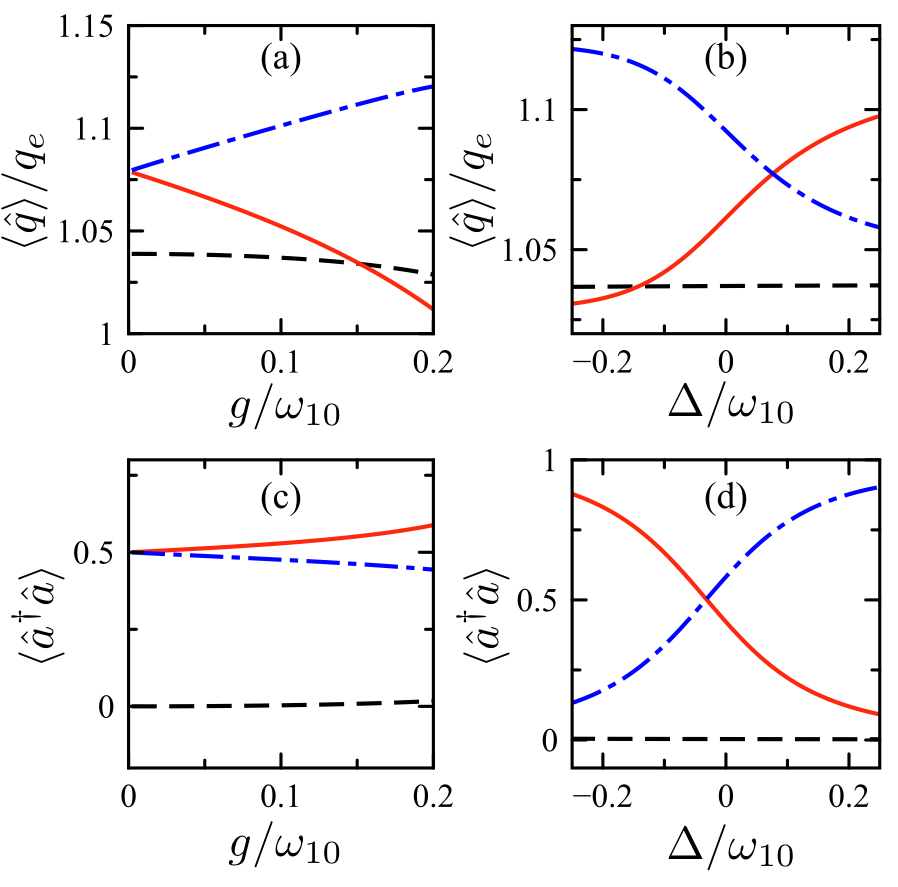}
\caption{Mean bond length $\langle \hat q\rangle$  and mean cavity photon number $\langle \hat a^\dagger \hat a \rangle $  as a function of coupling strength $g/\omega_{10}$ (a,c) and cavity detuning $\Delta$ (b,d), for the system ground state (dashed line), lower polariton (solid line) and upper polariton (dot-dashed line). We set $\Delta =0$ in panels a,c and $g/\omega_{10}=0.1$ in panels b,d. Energy is in units of $\omega_{10}$.}
\label{fig:Mean_pos_photnumb}
\end{figure}

\begin{figure}[t]
\includegraphics[width=0.48\textwidth]{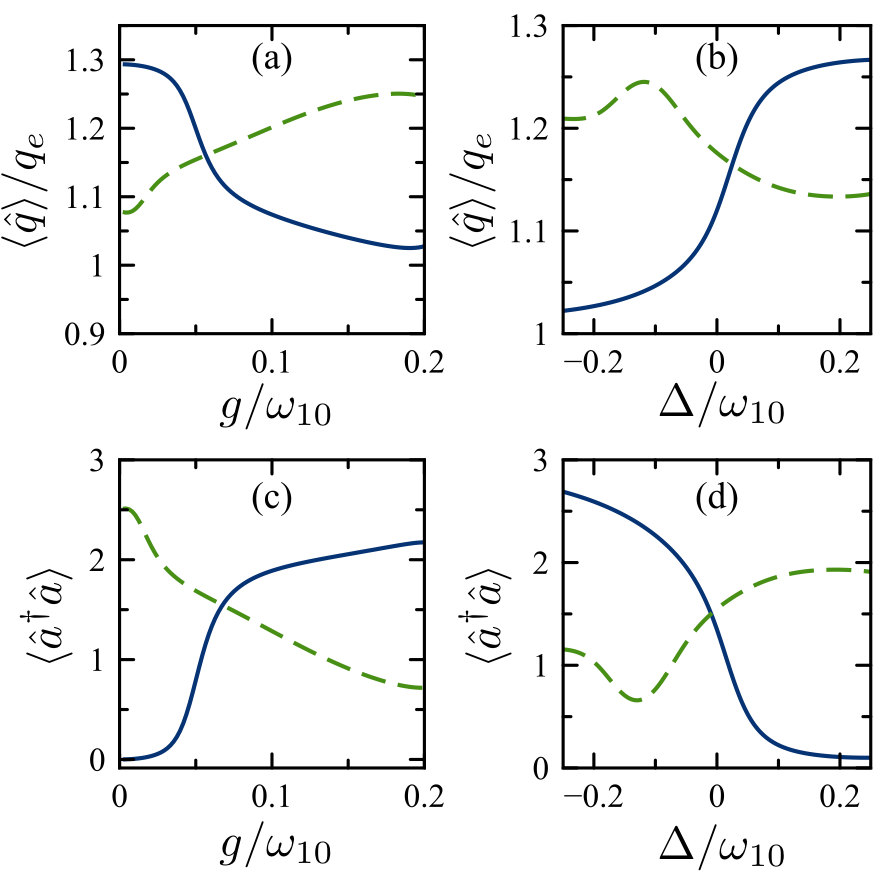}
\caption{Mean bond length $\langle \hat q\rangle$ and mean cavity photon number $\langle \hat a^\dagger \hat a \rangle $ as a function of coupling strength $g/\omega_{10}$ (a,c) and cavity detuning $\Delta$ (b,d), for excited polaritons $\ket{\Psi_6}$ (solid line) and $\ket{\Psi_8}$ (dashed line). We set $\Delta=0$ in panels a,c and $g/\omega_{10}=0.1$ in panels b,d. Energy is in units of $\omega_{10}$.}
\label{fig:Mean_pos_photnumb excited}
\end{figure}

In order to assess the contribution of each photon-number-dependent nuclear wave packet $\Phi_{j n}(q)$ on the $j$-th polariton eigenstate $\ket{\Psi_j}$, we show in Fig. \ref{fig:eigen_coeffs} the probability amplitudes $|c_{\nu n}|^2$ [see Eq. (\ref{eq:Psi_j expansion})] as a function of the coupling ratio $g/\omega_{10}$, for the excited polariton eigenstates $\ket{\Psi_6}$ and $\ket{\Psi_8}$. These two excited states tend asymptotically to the energy $E\approx 3\,\omega_{10}$ as $g\rightarrow 0$, and therefore can be expected to be mainly composed of uncoupled states $\ket{\nu}\ket{n}$ such that $\nu+n=3$, for resonant coupling. Figure \ref{fig:eigen_coeffs} shows that indeed occurs for $g/\omega_{10}\ll 1$. In this small coupling regime, the selected polariton eigenstates can be approximately written in the basis $\ket{\nu}\ket{n}$ as
\begin{equation}\label{eq:Psi6 low g}
    \ket{\Psi_6}\approx \ket{3}\ket{0}
\end{equation}
and
\begin{equation}\label{eq:Psi8 low g}
    \ket{\Psi_8}\approx a\ket{0}\ket{3} +b\ket{1}\ket{2},
\end{equation}
where $|a|^2\approx |b|^2=0.5$. As the coupling strength reaches the regime $g/\omega_{10}\sim 0.1$, the near resonant coupling between vibrational manifolds with higher photon numbers in Fig. \ref{fig:Coupling} leads to the emergence of wave function components with lower vibrational quantum numbers. For instance, for $g/\omega_{10}= 0.2$ the excited state $\ket{\Psi_6}$ is approximately given by
\begin{equation}\label{eq:Psi6 high g}
    \ket{\Psi_6}\approx a\ket{1}\ket{2} +b\ket{0}\ket{3} + c\ket{2}\ket{0} + d\ket{1}\ket{4}
\end{equation}
where $|a|^2\sim |b|^2>|c|^2 \gg |d|^2$. In other words, the state evolves from a bare Morse oscillator $\ket{\nu=3}$ in  vacuum [see Eq. (\ref{eq:Psi6 low g})], into a state with a lower mean vibrational excitation and higher mean photon number as $g/\omega_{10}$ grows. On the other hand, the state $\ket{\Psi_8}$ at $g/\omega_{10}= 0.2$ can be written as
\begin{equation}\label{eq:Psi8 high g}
    \ket{\Psi_8}\approx a\ket{3}\ket{0}+ b \ket{2}\ket{1}+ c\ket{0}\ket{3}+d \ket{1}\ket{4},
\end{equation}
where $|a|^2\approx 1/2> |b|^2>|c|^2 \gg |d|^2$, which also develops components with lower vibrational quanta and higher photon numbers in comparison with Eq. \ref{eq:Psi8 low g}. The emergence of uncoupled components with $\nu+n\neq 3$ in Eqs. (\ref{eq:Psi6 high g}) and (\ref{eq:Psi8 high g}) is a consequence of the counter-rotating terms in Eq. (\ref{eq:Rabi_Ham}). Although the results in Figs. \ref{fig:spacial_rep} and \ref{fig:eigen_coeffs} were obtained for specific polariton eigenstates, we find that they qualitatively describe the behavior of most excited polaritons $\ket{\Psi_j}$ with energies $E_j\gg \omega_{10}$, i.e., above the LP and UP frequency region.  

We can also understand the structure of vibrational polaritons in coordinate space and Fock space by analyzing the dependence of the mean bond distance $\langle \hat q\rangle$ and the mean photon number $\langle \hat a^\dagger \hat a \rangle$ with the coupling parameter $g$ and cavity detuning $\Delta$, for selected polariton eigenstates. In Fig. \ref{fig:Mean_pos_photnumb}, we compare the evolution of these observables with $g$ and $\Delta$ for the system ground state $\ket{\Psi_0}$ (GS), the lower polariton state $\ket{\Psi_1}$ and the upper polariton state $\ket{\Psi_2}$. In the regime $g/\omega_{10}\ll 1$, both LP and UP have the approximately the same bond length, given by
\begin{equation}\label{eq:LP UP length small g}
\langle\hat{q}\rangle\approx\frac{1}{2}\left(\langle 0|\hat{q}|0\rangle+\langle 1| \hat{q}| 1\rangle\right),
\end{equation}
with expectation value taken with respect to Morse eigenstates $\ket{\nu}$. 

\begin{figure*}[t]
\includegraphics[width=1\textwidth]{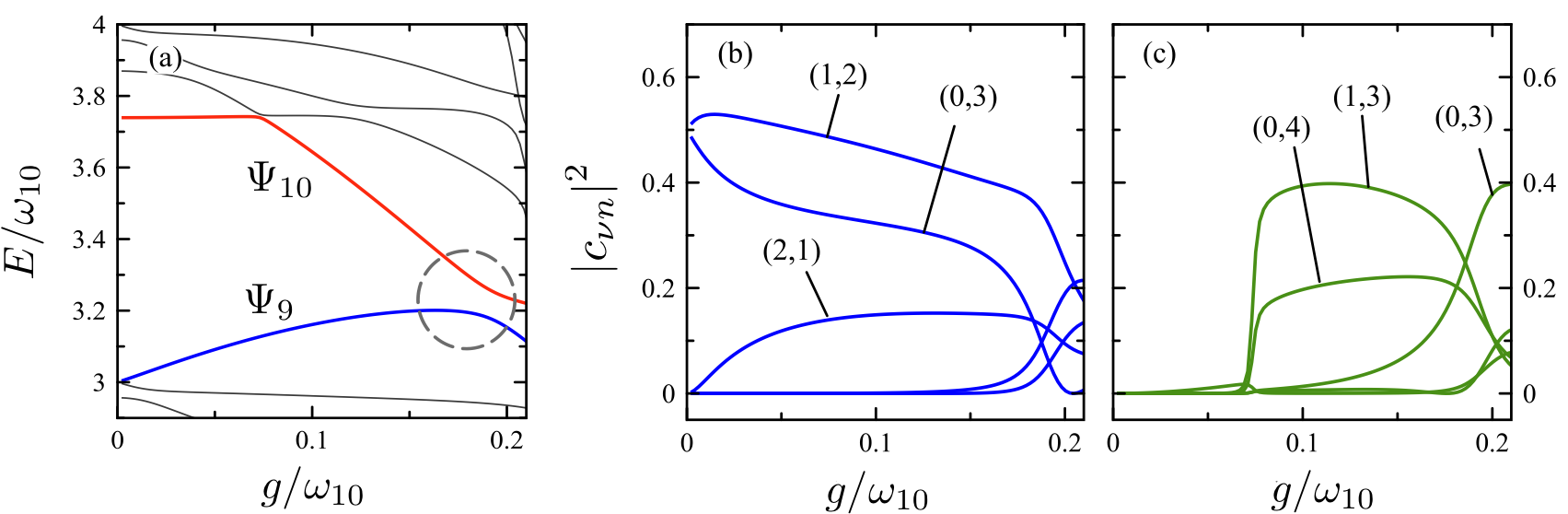}
\caption{(a) Spectral region with an avoided crossing (circled in grey) involving the excited polaritons $\ket{\Psi_9}$ (blue) and $\ket{\Psi_{10}}$ (red). (b) and (c) Main components of $\ket{\Psi_9}$ and $\ket{\Psi_{10}}$, respectively, in the uncoupled basis $\ket{\nu}\ket{n}$. Curves are labelled by the quantum numbers $(\nu,n)$. We set $\omega_c = \omega_{10}$.}
\label{fig:avoided crossing}
\end{figure*}

Figure \ref{fig:Mean_pos_photnumb}a shows that as the coupling strength increases, the bond length of the LP  decreases, reaching values even lower than the bond length of the Morse ground state $\ket{\nu=0}$. On the other hand, the bond length of the UP grows with increasing coupling strength. The value of $\langle\hat{q}\rangle$ for the UP is upper bounded by the bond length of the first excited Morse state $\ket{\nu=1}$. In other words, for resonant coupling the molecular bond in the LP state becomes stronger relative to the UP with increasing coupling strength, although both polariton states experience {\it bond hardening} in comparison with the Morse eigenstate $\ket{\nu=1}$, which is in the same energy region as LP and UP ($E/\omega_{10}\approx 1$). 

Bond hardening should be accompanied by the creation of virtual cavity photons, and bond softening by the decrease in the mean photon number. Figure \ref{fig:Mean_pos_photnumb}b shows that the GS, LP and UP states follow this behaviour as a function of $g/\omega_{10}$, for resonant coupling. We show in panels \ref{fig:Mean_pos_photnumb}b,d that for detuned cavities, the  compromise between bond strength and cavity photon occupation also holds. Within the range of system parameters considered, we find that this compromise also holds for higher excited vibrational polaritons, as Fig. \ref{fig:Mean_pos_photnumb excited} shows for states $\ket{\Psi_6}$ and $\ket{\Psi_8}$. 

Bond hardening of vibrational polaritons can be understood by recalling that an eigenstate $\ket{\Psi_j}$ in the vicinity of a bare Morse energy level $E_{\nu'}$ in general has non-vanishing components in the uncoupled basis $\ket{\nu}\ket{n}$ with $\nu<\nu'$ [see Eq. (\ref{eq:Psi_j expansion})]. These low-$\nu$ components contribute to the stabilization of the molecular bond even at high excitation energies.

\section{Energy crossings in the excited polariton manifold} \label{Energy crossings}

We discussed in Section \ref{Spec_sect} how the density of polariton levels increases with energy, ultimately due to the large number of near-degenerate uncoupled subspaces $\ket{\nu}\Ket{n}$ (see Fig. \ref{fig:Coupling}). Light-matter coupling leads to the formation of  closely-spaced polariton levels that can become quasi-degenerate at specific values of $g$ and $\Delta$. As the Hamiltonian parameters ($g,\Delta$) are tuned across the degeneracy point, the polariton levels may undergo true or avoided crossing. For a Hamiltonian like the quantum Rabi model for the qubit \cite{Braak2011,Wolf2013,Chen2012} and its multi-level generalizations \cite{Albert2012}, parity is a conserved quantity. Therefore polaritons in the quantum  Rabi model have well-defined parity and level crossings are analyzed in the usual way: states with opposite parity undergo true crossing under variation of a Hamiltonian parameter. In particular, the crossing of the ground state with the lower polariton state at $g/\omega_c=1$ marks the onset of the deep strong coupling regime \cite{Casanova2010,Forn-Diaz2018,Kockum2019}. 

Parity conservation in the quantum Rabi model ultimately emerges from the even character of the underlying microscopic Hamiltonians that describe the material system and the cavity field. The harmonic oscillator Hamiltonian that describes the cavity field is invariant under the transformation $\hat a\rightarrow -\hat a$, and therefore commutes with parity (as any harmonic oscillator Hamiltonian). For the material system, let $\hat q$ and $\hat p$ represent position and momentum operators in the material Hamiltonian $\hat{H}_{\rm M}$. Then polariton eigenstates of the coupled light-matter system would only have well-defined parity if $\hat H_{\rm M}$ is invariant under the parity transformation $\hat q\rightarrow -\hat q$ and $\hat p\rightarrow -\hat p$. The Morse potential in Eq. (\ref{eq:MO_pot}) is not invariant under the transformation $ q\rightarrow - q$ and $q_e\rightarrow -q_e$, and therefore breaks parity, which is the origin of vibrational overtones. Polariton eigenstates of the MLQR model therefore do not have well-defined parity. 

\begin{figure*}[t]
\includegraphics[width=1\textwidth]{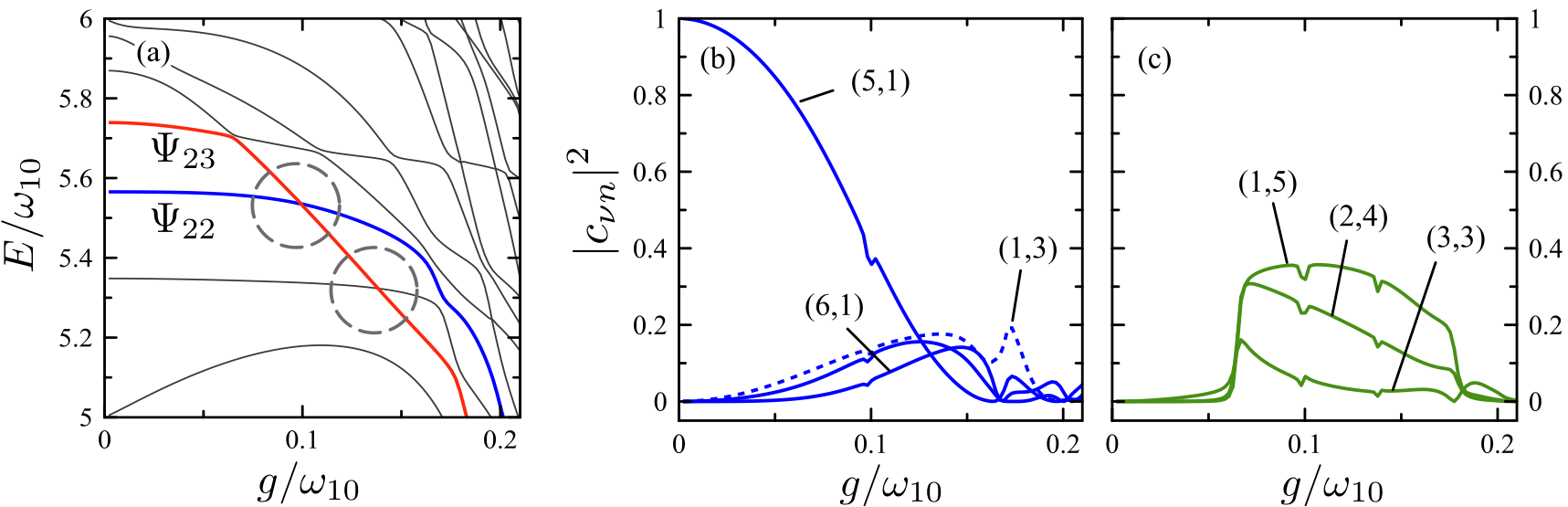}
\caption{(a) Spectral region with true crossings involving the excited polaritons $\ket{\Psi_{22}}$ (blue) $\ket{\Psi_{23}}$ (red), and $\ket{\Psi_{21}}$ and $\ket{\Psi_{23}}$ (both crossing regions circled in grey). (b) and (c) Main components of $\ket{\Psi_{22}}$ and $\ket{\Psi_{23}}$, respectively, in the uncoupled basis  $\ket{\nu}\ket{n}$. Curves are labelled by the quantum numbers $(\nu,n)$. We set $\omega_c=\omega_{10}$.}
\label{fig:true crossing}
\end{figure*}

Even parity is not a good quantum number, the vibrational polariton spectrum still exhibits true and avoided level crossings as the Hamiltonian parameters $g$ and $\Delta$ vary. We can track the origin of these crossings into an effective photonic parity selection rule imposed by the light-matter interaction term in the total Hamiltonian [Eq. (\ref{eq:Rabi_Ham})], which reads $\Delta n=\pm 1$.
For two near-degenerate polariton levels $E_j$ and $E_k$ ($k\neq j$), there will be a strong avoided crossing between them only if the largest probability amplitudes $c_{\nu n}$ of their wavefunctions $\ket{\Psi_j}$ and $\ket{\Psi_k}$ in the uncoupled basis $\ket{\nu}\ket{n}$, differ by one photon number [see Eq. \ref{eq:Psi_j expansion}]. Otherwise, the  levels will cross as the Hamiltonian is varied through the degeneracy. We show this explicitly in Figs. \ref{fig:avoided crossing} and \ref{fig:true crossing}, with examples of avoided and true crossings, respectively, in the excited polariton manifold. 

In Figure \ref{fig:avoided crossing}a we highlight an avoided crossing between excited polaritons $\ket{\Psi_{9}}$ and $\ket{\Psi_{10}}$, as the coupling strength is $g/\omega_{10}\approx 0.19$. Panels \ref{fig:avoided crossing}b and c show that the largest uncoupled components to the left of the avoided crossing are $\{\ket{1}\ket{2},\ket{0}\ket{3}\}$ for $\ket{\Psi_{9}}$ and $\{\ket{1}\ket{3},\ket{0}\ket{4}\}$ for $\ket{\Psi_{10}}$, which indeed differ by one photon number. Past the avoided crossing, the state $\ket{\Psi_{10}}$ dominantly acquires  $\ket{0}\ket{3}$ character.

In Fig. \ref{fig:true crossing}a we highlight a pair of level crossings as $g/\omega_{10}$ increases. For $g/\omega_{10}\simeq0.1$, the excited polariton states $\ket{\Psi_{22}}$ and $\ket{\Psi_{23}}$ undergo the first crossing. Fig. \ref{fig:true crossing}b shows that to the left of the crossing point, the largest uncoupled components of $\ket{\Psi_{22}}$  is $\ket{5}\ket{1}$, while for $\ket{\Psi_{23}}$ the largest components are $\{\ket{1}\ket{5},\ket{2}\ket{4}\}$. Since $\ket{\Psi_{22}}$ and $\ket{\Psi_{23}}$ thus predominantly satisfy $\Delta n > 1$, they do not interact via by the light-matter term as $g$ is varied across the degeneracy. Note that Fig. \ref{fig:true crossing}b,c the state does undergo a small change of character at the two crossing points in Fig. \ref{fig:avoided crossing}a. This occurs because states $\ket{\Psi_{22}}$ and $\ket{\Psi_{23}}$ do have uncoupled components that interact via the photonic selection rule $\Delta n\pm 1$, but their weight in the eigenfunction is comparatively small.

\section{Dipole self-energy} \label{Self_E_sect}

For simplicity, we have neglected the dipole self-energy term throughout. This term which arises via the transformation from minimal-coupling light-matter interaction to the multipolar interaction through the Power-Zineau-Wolley (PZW) transformation $U_{\rm PZW}= {\rm exp}[i\int d\mathbf{x} \mathbf{P}(\mathbf{x}) \cdot \mathbf{A}(\mathbf{x})]$
which eliminates the vector potential $\mathbf{A}(\mathbf{x})$ from the theory. Even though the multipolar formalism is non-covariant, it has been widely used to describe light-molecule interaction in the non-relativistic regime \cite{Craig-Thirunamachandran-book}. 

For light-matter coupling in which counter-rotating terms become important, it has been argued that the dipole self-energy contribution should be taken into account in order to describe polaritons correctly \cite{Kockum2019}. In molecular polariton problems, self-energy terms have been given ad-hoc model treatments in previous work \cite{George2016,Flick2017}. There, the dipole self-energy is considered to be proportional to the Rabi frequency $g$, which is in remarkable contrast with the PZW frame, in which a material Hamiltonian [see Eq. (\ref{eq:Hmol})] contains a dipole self-energy contribution even when light-matter coupling is perturbative. 

In principle, relating the macroscopic polarization density $\mathbf{P}(\mathbf{x})$ with the molecular electric dipole operator $\mathbf{d}(\mathbf{x}_0)$ at position $\mathbf{x}_0$ in the medium would require an {\it ab-initio} quantum electrodynamics formulation of field quantization in dispersive and absorptive media \cite{Knoll2001}, which is beyond the scope of our work. We ignore the contribution of the dielectric background and assume that for a single polar vibration the following ansatz holds
\begin{equation} \label{eq:H_self}
\hat H_{\rm self}=\int|\mathbf{P}(\mathbf{x})|^{2}dx \equiv \gamma\sum_{\nu}\bra{\nu}\hat{d}(q)\ket{\nu}^2 \ket{\nu}\bra{\nu},
\end{equation}
where $\ket{\nu}$ are the anharmonic vibrational eigenstates. In other words, we assume that self-energy leads to a state-dependent blue shift of every vibrational level. We can thus build a new total Hamiltonian $\hat{\mathcal{H}}'=\hat{\mathcal{H}}+\hat H_{\rm self}$, where $\mathcal{H}$ is the MLQR model from Eq. (\ref{eq:Rabi_Ham}). The parameter $\gamma$ is introduced to control the numerical magnitude of $\hat H_{\rm self}$ relative to $\hat{\mathcal{H}}$.

\begin{figure}[t]
\includegraphics[width=0.47\textwidth]{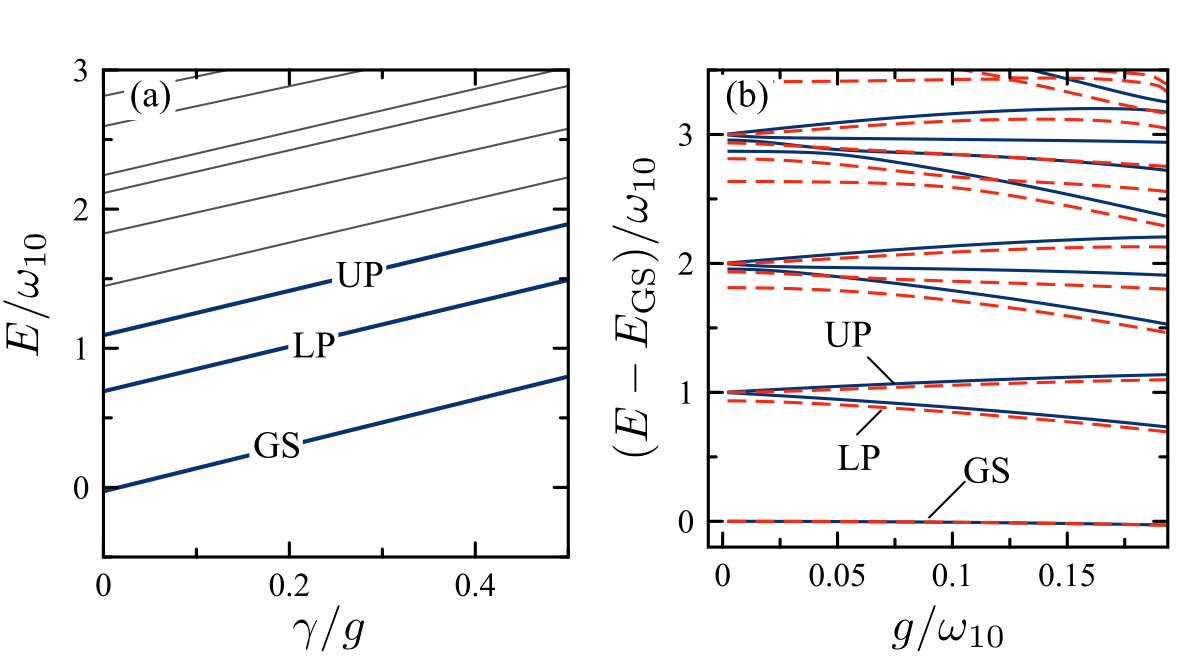}
\caption{(a) Vibrational polariton spectrum as a function of the dipole self-energy parameter $\gamma$, for $g=0.2\omega_{10}$ and $\omega_c=\omega_{10}$. (b) Comparison of the polariton spectrum as a function of the coupling strength $g/\omega_{10}$ using the MLQR model without (solid line) and with (dashed line)  the dipole self-energy term with $\gamma = 0.15$. Energy is given relative to the ground state (GS), in units of $\omega_{10}$.}
\label{fig:self-energy}
\end{figure}

In Fig. \ref{fig:self-energy}a, we show the polariton spectrum with increasing $\gamma$, for $g$ and $\Delta$ fixed. As expected \cite{George2016,Flick2017}, $\hat H_{\rm self}$ only results in a state-dependent positive energy shift in the polariton levels. Once the dipole-shifted ground state energy $E_{\rm GS}$ is subtracted from the energies of $\mathcal{H}'$, Fig. \ref{fig:self-energy}b shows that the energy spectrum of $\mathcal{H}'$ has the same qualitative behavior with increasing $g$ as the spectrum of $\hat{\mathcal{H}}$. Quantitative deviations from the MLQR spectrum due to self-energy  become important for relative energies $(E-E_{\rm GS})\gtrsim 2\,\omega_{10}$, when the magnitude of the parameter $\gamma$ is comparable with the coupling ratio $g/\omega_{10}$.

\section{Conclusion and Outlook} \label{Discussion and Outlook}

In order to understand the microscopic behaviour of an individual anharmonic molecular vibration coupled to a single  infrared cavity mode, we introduce and analyze the multi-level quantum Rabi (MLQR) model of vibrational polaritons [Eq. (\ref{eq:Rabi_Ham})]. We derive the model Hamiltonian starting from the exact anharmonic solutions of a free-space Morse oscillator, and treat light-matter interaction within the Power-Zineau-Wolley multipolar framework \cite{Craig-Thirunamachandran-book}, which includes the dipole self-energy. The model takes into account counter-rotating terms in the light-matter coupling and allows the analysis of vibrational polaritons both in Hilbert space and nuclear coordinate space. Phase-space representations of the photon state follow directly from the QED formulation of the model \cite{Carmichael-book1}. Such phase-space analysis would be closely related to previous coordinate-only treatments of photon-nuclei coupling \cite{Flick2017,Feist2018,Triana2018}, although a systematic comparison has yet to be done.  

The model is consistent with previous work based on few-level vibrational systems \cite{Ribeiro2018}, and therefore is also able to describe the spectral features observed in linear and nonlinear transmission spectroscopy \cite{muallem2016strong,Xiang2018,Dunkelberger2016}, which due to the relatively weak intensities involved, can only probe up to the second excited polariton triplet around $E\approx 2\,\omega_{10}$, where $\omega_{10}$ is the fundamental vibration frequency. Few-level vibrational truncations are however unable to capture the dense and complex polariton level structure predicted by the MLQR model at energies $E\gtrsim 3\,\omega_{10}$. The system Hamiltonian allows the emergence of an ensemble of avoided and true crossings as the Rabi frequency $g$ and cavity detuning $\Delta$ are tuned. The density of these level crossings increases with energy. These crossings are governed by a pseudo-parity selection rule in the photonic degree of freedom (details in Sec. \ref{Energy crossings}).

The nuclear coordinate analysis of vibrational polaritons within the MLQR model unveils a few general trends accross the entire energy spectrum. First, it is no longer possible to define a unique bond dissociation energy in an infrared cavity as is commonly done in free space. The dissociation energy depends on the quantum state of the cavity field. Second, within any given energy range $E_j+\Delta E$, it is always possible to find a vibrational polariton eigenstate with small mean photon number $\langle \hat a^\dagger \hat a\rangle $ and large mean bond distance $\langle \hat q\rangle$, and vice-versa. Third, the bond distance $\langle \hat q\rangle$ of an arbitrary vibrational polariton state with energy $E_j$, never exceeds the bond length of a free-space Morse eigenstate $\ket{\nu}$ with similar energy ($E_\nu\approx E_j$). In other words, the formation of vibrational polaritons inside the cavity leads to a type of {\it bond-hardening} effect that may have consequences in the reactivity of chemical bonds.

The generalization of the multi-level quantum Rabi model developed here to the many-molecule and multi-mode scenarios is straightforward. Since it is formulated in the energy eigenbasis, treating the dissipative  dynamics of vibrational polaritons due to cavity photon decay and vibrational relaxation is also straightforward to formulate within a Markovian approach \cite{Breuer-book}. The dynamics of vibrational polaritons in the many-body regime has been previously discussed in Refs.  \cite{DelPino2015,Strashko2016}, using truncated vibrational subspaces. The main qualitatively new effect that the many-body system introduces to the problem, is the formation of collective molecular states that are not symmetric with respect to particle permutations. These so-called ``dark exciton states" \cite{Litinskaya2004} arise naturally from state classification by permutation symmetry in the Hilbert space of the Dicke model \cite{Garraway2011,Kirton2019}. It has been shown originally within a quasi-particle approach for systems with macroscopic translational invariance \cite{Litinskaya2006-disorder}, and later using a cavity QED approach \cite{Herrera2016,Herrera2017-PRA,Herrera2017-PRL,Herrera2017-review}, that  totally-symmetric and non-symmetric collective molecular states can strongly admix due to  ever-present inhomogeneous broadening of molecular energy levels,  inhomogeneities in the light-matter interaction energy across the medium, or  any local coherent term such as intramolecular electron-vibration coupling (in the case of electronic strong coupling \cite{Herrera2017-review}). In general,  the role of quasi-dark collective states in determining the rate of chemical reactions and also spectroscopic signals of vibrational polaritons is yet to  be fully understood.

\acknowledgements
We thank Guillermo Romero, Blake Simpkins and Jeffrey Owrutsky for discussions. This work is supported by CONICYT through the Proyecto REDES ETAPA INICIAL, Convocatoria 2017 no. REDI 170423, FONDECYT Regular No. 1181743, and also thank support by Iniciativa Cient\'{i}fica Milenio (ICM) through the Millennium Institute for Research in Optics (MIRO).

\bibliographystyle{unsrt}
\bibliography{morse-polariton}

\end{document}